# FINANCE 4.0: DESIGN PRINCIPLES FOR A VALUE-SENSITIVE CRYPTOECONOMIC SYSTEM TO ADDRESS SUSTAINABILITY

*Research Paper*

Mark C. Ballandies, ETH Zurich, Zurich, Switzerland, mark.ballandies@gess.ethz.ch

Marcus M. Dapp, ETH Zurich, Zurich, Switzerland, marcus.dapp@ethz.ch

Benjamin A. Degenhart, ETH Zurich, Zurich, Switzerland, b.degenhart.work@gmail.com

Dirk Helbing, ETH Zurich, Zurich, Switzerland, dirk.helbing@gess.ethz.ch

## Abstract

Cryptoeconomic systems derive their power but can not be controlled by the underlying software systems and the rules they enshrine. This adds a level of complexity to the software design process. At the same time, such systems, when designed with human values in mind, offer new approaches to tackle sustainability challenges, that are plagued by commons dilemmas and negative external effects caused by a one-dimensional monetary system.

This paper proposes a design science research methodology with value-sensitive design methods to derive design principles for a value-sensitive socio-ecological cryptoeconomic system that incentivizes actions toward sustainability via multi-dimensional token incentives. These design principles are implemented in a software that is validated in user studies that demonstrate its relevance, usability and impact. Our findings provide new insights on designing cryptoeconomic systems. Moreover, the identified design principles for a value-sensitive socio-ecological financial system indicate opportunities for new research directions and business innovations.

## 1   Introduction

Many environmental problems represent a commons dilemma, a "social situation in which a collective cost or risk is generated via the combined negative externalities of numerous individuals" (Vlek and Steg 2007, 9). The more individuals are involved and the smaller the individual negative external effect is, the harder the management of a common dilemma becomes. Many current policy approaches deal with the symptoms of environmental degradation, rather than providing "incentives for sustainable practices" (Fischer et al. 2012, 2). Fischer et al. argue that civil society institutions should be strengthened, and sustainability is in need of a "social avalanche of unprecedented proportions" (158). However, existing institutions and consumption patterns reinforce disincentives for sustainable actions of citizens (Fischer et al. 2012, 159; Vlek and Steg 2007, 8). On the other hand, recent developments in distributed ledger technology (DLT) draw together cryptography/computer science, complex systems science, and information systems (IS) to explore "cryptoeconomic systems" (CES) incentivizing human behaviour (Voshmgir and Zargham 2019a). Therefore, this paper responds to the call by Rossi et al. (2019) for an inclusive research agenda for DLT research in IS. Tackling the commons dilemma by CES design, this paper addresses the following research questions.

*Research Question 1: What role can IS play in shaping incentives for sustainable behavior?* This paper outlines and demonstrates how IS can nurture a new generation of hybrid techno-socioeconomic systems. The advent of distributed ledger technology (DLT) allows one to develop software artifacts whose storage and computation components may become immutable against manipulation over time, if adequate properties of the enabled CES emerge. Then, users of such a cryptoeconomic network can trust that the rules enforced by the system are not prone to unilateral manipulation (Ballandies, Dapp, and Pournaras 2021, Forthcoming).

Fair and transparent governance is a key challenge in the context of sustainability as the 'tragedy of the commons' is characterized by information asymmetries, conflicts of interest, etc. (Hardin 1968). Although solutions have been proposed (like Ostrom's (1990) principles for common pool resource management), they do not easily scale to thousands or even millions of people (Ostrom et al. 1999; Ostrom 2008). In contrast to solely human-controlled settings, cryptoeconomics aims to design complex systems with in-built rule sets that are self-(re)-enforcing. This makes it even more important to follow a value-sensitive design paradigm for the underlying software system (Friedman, Kahn, and Boring 2008; Huldtgren 2015). Thus, a second contribution of this paper is to report a consistent methodology for the design and implementation of such a value-sensitive system by answering

*Research Question 2: How can a cryptoeconomic system be designed and implemented in a value-sensitive way that allows one to study the impact of cryptoeconomic tokens on human behavior?* Using design science research and value-sensitive design, a framework and system is developed that enables researchers and practitioners to design, implement and test CES: Inspired by nature, the outlined "Finance 4.0" (FIN4) system aims to address one of the key problems in the sustainability debate: The handling of negative externalities. The approach aims to "extend the economic system itself by systematically including externalities and making them tradable on markets" (Dapp 2019, 157). Moreover, the system also supports scientific "in-vitro" experiments in behavioral economics by enabling a large spectrum of token economies to be designed and "played out" by human and/or machine users. A particular novelty from an economic standpoint is the use of a multi-dimensional incentive system that uses a number of tokens to capture different value dimensions of a good or service (Kleineberg and Helbing 2016).

*Reserch Question 3: What are the design principles for a cryptoeconomic system that enable a socio-ecological finance system?* Design knowledge for cryptoeconomic systems in general is still relatively scarce (Meier et al. 2021; Beck et al. 2016; Lockl et al. 2020). Thus, the third contribution of this paper are key design principles for a software system (called FIN4 Xplorer, FIN4X) that enables a multi-dimensional socio-ecological financial system (called FIN4). The principles are derived by assigning and translating the list of stakeholder values into a multi-layered structure that represents the interactions between the FIN4X system to be built and the cryptoeconomic system FIN4 enabled by it. The principles instantiated in the FIN4 system are tested in user studies.

Broken down to the research agenda proposed by Rossi et al. (2019), the individual contributions are as follows: To *Behavioral IS research*, the paper contributes (a) a validated constraint configurator that facilitates research on the impact of blockchain protocols on user behavior by enabling the creation of cryptoeconomic incentive systems, as well as (b) a governance mechanism of a DLT system applying token-curated registries (TCR). To *Design Science Research*, it offers (a) design principles for a socio-ecological financial system; (b) a design and implementation methodology for DLT systems that combines a value-sensitive design methodology (Friedman, Kahn, and Boring 2008) with the design-evaluate-construct-evaluate pattern (Sonnenberg and Vom Brocke 2012). This novel design methodology can mitigate unintended side effects of new blockchain implementations by considering values explicitly during design; (c) best-practices for comprehensively illustrating the design of cryptoeconomic IS systems. Finally, to *Economics of IS Research*, it contributes (a) the conceptualisation and implementation of a universal value creation and capturing mechanism for DLT systems -- the token creator; (b) the Conceptualisation and implementation of an external source of value for tokens; and (c) methods for capturing and representing real-world actions on-chain in a trustworthy way. The paper is structured as follows. Section 2 discusses literature that contributes to the emerging domain of cryptoeconomics (also called token engineering). Section 3 answers RQ2 by describing the design science research methodology and evaluation patterns applied. Section 4 answers RQ1 and RQ3 by describing the process of creating the FIN4 system alongside four research activities including the design for values process. Section 5 offers a summary of findings followed by conclusions in Section 6.

## 2 Related Work: The emerging research field of Cryptoeconomics

This section draws together selected literature from different disciplines -- computer science, complex systems science, and information systems -- to show how they each can contribute to the formation of the emerging field of research called 'cryptoeconomics'. Computer science, in particular cryptography, was the first discipline to address cryptocurrencies, blockchains, and distributed ledgers. Contributions come from cryptography (Böhme et al. 2014; Bonneau and Heninger 2020; Bonneau et al. 2015), distributed systems (Wattenhofer 2016) and software engineering (Antonopoulos 2017; Antonopoulos and Wood 2018). One of the earliest contributions in this field defines *cryptoeconomics* as building systems that "use cryptography to prove properties about messages that happened in the past" and that use "economic incentives defined inside the system to encourage desired properties to hold into the future" (Buterin 2017). Voshmgir and Zargham describe *cryptoeconomic systems (CES)* from a complex system perspective by illustrating it via three interconnected networks: "(i) the computation and communication network comprised of nodes that leverage a peer-to-peer protocol to validate transactions by mining new blocks, (ii) the financial network comprised of Bitcoin addresses which may sign transactions and transfer funds, and (iii) the off-chain socioeconomic network representing people and organizations that control the tokens in the financial network and operate those nodes in the computation and communication network" (Voshmgir and Zargham 2019b, 7). Such a complex systems perspective is useful when researching CES because it reveals an important distinction: the underlying *software system* is not identical to the *CES*. The latter may emerge "on top" only if the former has been adequately designed in terms of economic incentives and constraints. In particular, typical control attempts in these systems "are destined to fail" (Helbing 2008, 13) because of the *emergent properties in complex systems* such as non-linearity of cause and effect, self-organization, phase transitions, cascading effects, and self-organized criticality - to which complexity science can provide insights and tools (Helbing 2008, chap. 1). In particular, agent-based simulations and experiments are utilized to anticipate and enable social self-organization (Helbing 2012). For a broad overview into systems thinking see (Mainzer 2007). In Information Systems, CES affect *digital infrastructures* research (Tilson, Lyytinen, and Sørensen 2010): CES shift the boundaries between the opposing logics of centralized and distributed control ("paradox of control") in the evolution of digital infrastructures, as well as the opposing logics of stability and flexibility ("paradox of change") across the infrastructure layers as described by Tilson, Lyytinen, and Sørensen (Tilson, Lyytinen, and Sørensen 2010, 753f). Two examples: (i) CES push the boundary towards decentralized control by aiming at governance mechanisms that assume anonymous peers in a non-hierarchical setting as opposed to traditional institutions. (ii) They also take the paradox of change to extremes: some maximize flexibility taking into account even fundamental changes like replacing the consensus algorithm (e.g. Ethereum), while others aim to maximize stability by "ossifying" the protocol (e.g. Bitcoin). Ballandies et al offer a taxonomy of DLT system (Ballandies et al. 2021, Forthcoming), while Tönnissen et al. offer a taxonomy of blockchain-based business models (Tönnissen, Beinke and Teuteberg 2020).

In 2019, two larger initial contributions came from IS. First, an extensive 2-volume book on business transformation through blockchain with a wide range of contributions was published (Treiblmaier and Beck 2019), also including a first introduction to cryptoeconomics in the context of sustainability (Dapp 2019). Second, a special edition of JAIS discussed opportunities and challenges of blockchain technology on issues like self-organization of blockchains, privacy in IoT networks, and business models (Andersen and Bogusz 2019; Chanson et al. 2019; Chong et al. 2019). Moreover, a contribution in ICIS investigated the performance of token-based incentives in prediction markets via agent-based modeling (Hülsemann and Tumasjan 2019).

This paper contributes to this cryptoeconomic research in IS by answering to the call by Rossi et al. (2019) for a dedicated *blockchatin research agenda*. In particular, it addresses questions related to behavioral (individual, group, and organizational), design science, and IS economics research on blockchain. In addition, by choosing the financial system as an application domain, this paper contributes to the *control of systemic risks in global networks*, which was identified as the top-ranked 'Grand Challenge' for IS research in a recent survey (Mertens and Barbian 2015, 394). The crisis in 2008 and the lockdowns in 2020 have shown that the financial system is a prime example of a global network exhibiting systemic risks humans need to address.

## 3 Methodology

This paper follows an interactive Design Science Research Methodology (Hevner and Chatterjee 2010; Hevner et al. 2004; Vom Brocke et al. 2020; Gregor and Hevner 2013) to address the above formulated research questions (Section 1). For this, it applies general DSR evaluation patterns (Vaishnavi and Kuechler 2015) in a cyclic DSR process as introduced by Sonnenberg et al. (2012) to identify value-sensitive design principles (Gregor et al. 2020) of a CES and implement them in a validated software artefact. By using this cyclic DSR process (Sonnenberg and Vom Brocke 2012), the research is sequenced into four, individually evaluated, activities: problem identification, design, construction, and use. During the problem identification and design activities, design workshops as a research method (Ørngreen and Levinsen 2017) are used to derive the values and design principles of the system, which is a common method in information systems (Peffers et al. 2007). These workshops are illustrated in greater detail in Section 3.1.The value-sensitive design approach is introduced in Section 3.2 and the evaluation patterns in Section 3.3.

*Design Workshops:* Throughout the first two activities, design workshops (Table 1) calibrated, evaluated and provided knowledge for the analysis of the research questions and design of the software artifact. These workshops were conducted during the first two activities of the DSR process (Identify Problem, Design) and are classified as structured, semi-structured, or open. Structured focus group meetings follow a strict methodology, whereas open meetings are characterized by an open discussion, during which the authors take notes. Semi-structured meetings let every participating institution present their view on a particular topic followed by an open discussion. In total, participants from 34 institutions discussed 13 topics in 22 focus group sessions.

*Value-sensitive Design (VSD)* is "a theoretically grounded approach to the design of technology that accounts for human values in a principled and comprehensive manner throughout the design process." (Friedman, Kahn, and Boring 2008). This research followed VSD by (i) identifying the underlying values of the system and its stakeholders during the stakeholder mapping; (ii) considering the identified values in requirements analysis and (iii) validating the systems according to the values in a dedicated workshop on ethics (23.05.2019, Table 1).

*Evaluations:* Three evaluation patterns -- demonstrations, case studies, and prototyping -- were utilized (Sonnenberg and Vom Brocke 2012). In *demonstrations*, focus groups interacted with the software, built upon it and provided user feedback. Five types of focus groups were Associate Editor's Meta-Review to Track Chairs# used: (i) dedicated Forums (2/110)[1] illustrated the prototype to an interested audience and enabled them to interact with the system. This format spanned one or two days and involved a large number of stakeholders that obtained an in depth introduction to the artifact; (ii) Hackathons (4/65), during which teams of participants used concepts and tools of the software artifact to improve the artifact and to build own prototypes; (iii) Presentations (6/180) illustrated the core concepts of the artifact's functionalities in non-interactive live demos; (iv) Workshops (6/125) provided a live demonstration of the software artifact and let participants interact with the software; (v) Webinars (4/44) were virtually conducted Workshops. In total, 524 participants in 22 focus groups provided feedback to different versions of the software artifact. In *case studies*, the use of the software instance in communities and organizations was evaluated. For this, participants tested the system in a real-world environment; and evaluated the acceptance and usability with a study (N=8)[2]. An online questionnaire based on the 'Unified Theory of Acceptance and Use of Technology' (Venkatesh, Thong, and Xu 2012; Zhou 2012) was used to evaluate user acceptance. The results were used to validate the software in Associate Editor's Meta-Review to Track Chairs#stance. Across the phases of the project, Agile Software development for *prototyping* (Paetsch, Eberlein, and Maurer 2003) was applied to implement the software instance. Characteristic for this iterative approach is the early involvement of users, which is realized through the involvement of potential users in the focus groups.

---

[1] (X/Y) means the number of sessions X and participants Y per focus group type.
[2] The results are preliminary. Due to the ongoing pandemic situation in 2020, tw Associate Editor's Meta-Review to Track Chairs#o field experiments involving 50+ participants had to be postponed.

| Dates | Activity | Type | Participating institutions | Name/ Topic |
|---|---|---|---|---|
| 09.05.2017 | Identify Problem | semi-structured | ETH Zürich, ValidityLabs, PwC, The Humanized Internet (presenters) | E-Identity |
| 20.09.2017 | Identify Problem | semi-structured | Bafin, ECB, Frankfurt School of Finance, KfW, Fortschrittszentrum, Bundesbank | Designing sus. economic system using ICT |
| 02.10.2017 05.10.2017 | Identify Problem | open | ETH Zurich | Specification |
| 24.04.2018 | Design | structured | ETH Zurich, UNICEF, FuturICT 2.0, TU Munich | Stakeholder mapping |
| 02.05.2018 16.05.2018 22.05.2018 | Design | structured | ETH Zurich, UNICEF, FuturICT 2.0, TU Munich | Requirements Engineering |
| 17.05.2018 | Design | open | ETH Zurich, UNICEF, TU Munich | Identity Management |
| 27.08.2018 | Design | open | ETH Zürich, CNR, Politecnico, ULB, University of Tartu, RTU, Tallinn University of Technology | Managing complex, global, socially interactive systems |
| 22.11.2018 | Design | semi-structure Associate Editor's Meta-Review to Track Chairs | Airalab RU, Netis SI, Agoora CA/CH, ETH Zurich, | Proving/Oracles |
| 06.12.2018 | Design | semi-structured | ETH Zurich, Astratum, Research Institute for Cryptoeconomics (WU), TU Munich, dwf, Berlin Inno Ventures | Cryptoeconomics |
| 04.04.2019 30.04.2019 08.05.2019 20.05.2019 | Design | open | ETH Zurich, University of Bremen, ETH library lab, TU Munich | Token Curation |
| 23.05.2019 | Design | structured | Ethix, ETH Zurich | Ethics |
| 18.11.2019 | Design | semi-structured | Kleros; ETH Zurich; Procivis; CNR Rome; RIAT; TU Munich; SIX Group | Governance |
| 28.01.2020 03.03.2020 13.03.2020 16.04.2020 | Design | open | ETH Zurich, CNR Rome; | Reputation, Agents-based modeling |

*Table 1  Design workshops held to specify the incorporated values, design principles and final layout of the FIN4 system.*

## 4 The Finance 4.0 system: Multi-dimensional incentivization

In the following, we illustrate the findings and artifacts obtained at each activity of the cyclic DSR process (Sonnenberg and Vom Brocke 2012).

## 4.1  Problem Identification and Solution

Non-sustainability has be found to be one of the greatest challenges humanity is facing at the beginning of the 21st century (Dapp 2019). In the past, it was tried to solve sustainability issues by means of laws and regulation (Rogelj et al. 2016). By now, however, we can say it has not solved the world's problems on time (Seele et al. 2019). We, therefore, need a new approach to tackle the challenge. Here, we are proposing a bio-inspired approach (Dapp, Helbing, and Klauser 2021b, Forthcoming). Ecosystems are very impressive in terms of their logistics and recycling (Helbing et al. 2009). Nature has already managed to build something like a circular economy, i.e. closed cycles of material flows. It did not get there by regulation and optimisation though, but by (co-)evolution -- a principle, which is based on the self-organisation of complex systems.

Optimisation, in contrast, which is often used in economics, tends to be based on a one-dimensional goal function and, therefore, to oversimplify the needs of complex systems. In particular, it often neglects other, non-aligned goals. Of course, there are also methods for multi-objective optimization (Marler and Arora 2004), but co-evolution as we find it in nature seems to work differently, based on mutation, selection, and multiple feedback loops (Grund et al. 2013, Helbing 2013). Using such principles underlying self-organization, complex systems may improve over time in a variety of aspects. A one-dimensional incentive system such as money cannot accomplish this task in the same way as multi-dimensional incentive systems can do. Therefore, we are proposing to build a multi-dimensional money system. How would such a socio-ecological finance system have to work? It would have to be able to incentivize differentiated activities considering various different externalities. For this purpose, we couple measurements (of quantities such as $CO_2$, noise, waste and poisons as well as positive effects) with new kinds of currencies. This is done by using measurements with Internet of Things sensors to define new kinds of currencies and incentives, using blockchain technology (Dapp, Helbing, and Klauser 2021a, Forthcoming). In this way, a multi-dimensional real-time feedback and coordination system can be built. It will enable a better management of complex systems, and even the design of systems capable of self-organisation (Helbing 2016). These new incentives, which may try to incentivize the reduction of noise, $CO_2$, and waste, or the planting of trees, to give just a few examples, can introduce new forces into our socio-economic system, thereby promoting sustainable and environmental-friendly behavior. Over time, based on the co-evolutionary process kicked off by these multiple incentives, this will promote the emergence of a circular and sharing economy. For this reason, we are proposing a CES with a variety of tokens, coupled to measurements or other validation procedures. The approach taken is one of value-sensitive design (Friedman and Hendry 2019; Van den Hoven, Vermaas, and Van de Poel 2015; Shahriari and Shahriari 2017), i.e. design that builds in constitutional, cultural, and other values. Such an approach can support human rights and environmental goals, while promoting economically efficient solutions.

## 4.2  Design of Software

During the design activity, the key values (Section 4.4.1) and design principles (Section 4.4.2) of a socio-ecological multi-dimensional finance system are identified that are utilized in the design of the proposed CES (Section 4.4.3)

### 4.2.1  Value-sensitive design

During a structured focus group meeting, a stakeholder map (Clayton 2014) has been created by the participants and subsequently published (Rössner, Dapp, and Fach 2018). Three stakeholder clusters were identified (cf. Table 2). Cluster I consists of stakeholders that are less interested in and have medium influence on the FIN4 system (private companies, central banks, IoT/hardware providers, Cryptocurrency owners). Cluster II consists of stakeholders that have a high interest in the FIN4 system, while having medium influence on its success (academic and governmental institutions). For them, regular information about the system development (unidirectional communication) is adequate. Cluster III consists of stakeholders that have both, high interest in the system and a high influence on its success, and thus should be addressed primarily: Individual Users and Communities. With them, intense bidirectional communication and exchange is required particularly to consider their perspectives and values during design and implementation (e.g., focus groups). Finally, no parties were identified who actively oppose the system. Next, the relevant values for each stakeholder group have been identified from the value-sensitive design literature focusing on information and communication technology (Huldtgren 2014; Friedman, Kahn, and Boring 2008; Van de Poel 2015; Harbers and Neerincx 2017; Barn, Barn, and Primiero 2014). These values are illustrated in Table 2. The importance indicates that stakeholders, particularly from Cluster III, value resilience, autonomy, inclusiveness, and trust the most, and give sustainability, privacy, and efficiency medium importance. All values have been translated into system requirements, for details cf. Rössner, Dapp, and Fach (2018).[3]

| ID | Cl. | Stakeholder | Tru | Aut | Inc | Res | Sus | Pri | Eff |
|---|---|---|---|---|---|---|---|---|---|
| 1 | I | Private Companies | | x | | | | x | x |
| 2 | I | Cryptocurrency owners | x | x | | x | | x | |
| 3 | I | IoT/Hardware provider | x | | | | | | x |
| 4 | I | Central Bank experts | | | | | | | x |
| 5 | II | FuturICT 2.0 academic partners | x | x | x | | x | | |
| 6 | II | World Bank/ UN Agencies/ WWF | x | | x | x | x | x | x |
| 7 | II | Universities | x | x | | | | x | |
| 8 | II | Cities/ Regions | x | x | x | | x | x | x |
| 9 | III | FIN4 technology partners | x | x | x | x | x | | |
| 10 | III | Complementary Currency groups | x | x | x | x | | | x |
| 11 | III | Individual users and communities | x | x | x | x | x | x | |
| | | *Weighting all stakeholders* | *9* | *8* | *6* | *5* | *6* | *6* | *6* |
| | | *Weighting cluster III stakeholders* | *3* | *3* | *3* | *3* | *2* | *1* | *1* |
| | | ***Importance rating (high/medium)*** | ***H*** | ***H*** | ***H*** | ***H*** | ***M*** | ***M*** | ***M*** |
| | | *Number of requirements addressing value* | *8* | *>8* | *>8* | *4* | *1* | *2* | *3* |

*Table 2    Mapping of stakeholder clusters as identified in the stakeholder map (Section 4.2.1), stakeholders, values and requirements. The values are: (Tru)st, (Aut)onomy, (Inc)lusiveness, (Res)ilience, (Sus) tainability, (Pri)vacy, and (Eff)iciency.*

### 4.2.2 Design Principles for FIN4

During the series of design workshops spanning three years (cf. TABLE 1), the stakeholder values were gradually translated into design principles that informed the cryptoeconomic design of FIN4. These design principles represent the translation of stakeholder values into aspects of the CES to be built. The value *Trustworthiness* is taken into account (i) by making the software code *transparent* as open source, and (ii) by using a *public auditable* distributed ledger. The value *Autonomy* is taken into account (i) by choosing a *permissionless distributed ledger and open source* smart contract engine, and (ii) by allowing participation via a registration-less *opt-in* system. *Inclusiveness* is taken into account (i) by *enabling every user to participate* in creating, obtaining, and curating tokens, verifying proofs, and in democratic decision-making, and (ii) by a *direct-democratic* token curation mechanism. *Resilience* is taken into account (i) by using a distributed-ledger-type distributed system for data storage and computation, and (ii) by enabling a *multi-dimensional* token-based incentive system. In addition: *sustainability* as a value is taken into account by connecting token issuance to the *requirement of positive action* that is verifiable, and thus takes care of the internalization of negative economic externalities. *Privacy* is taken into account by allowing use under *pseudonymity* (no clear names required) and to build identity via reputation only. Finally, *efficiency* is taken into account by automating all interaction patterns via a decentralized application. Altogether, eight design principles cater for the four values with a high importance rating and another three cater for the three values of medium importance (Table 4).

---

[3]    Note that sustainability, while one of the core missions of the entire FIN4 system, is not a software design requirement but a property of the cryptoeconomic system *emerging,* if the rule sets have been defined adequately.

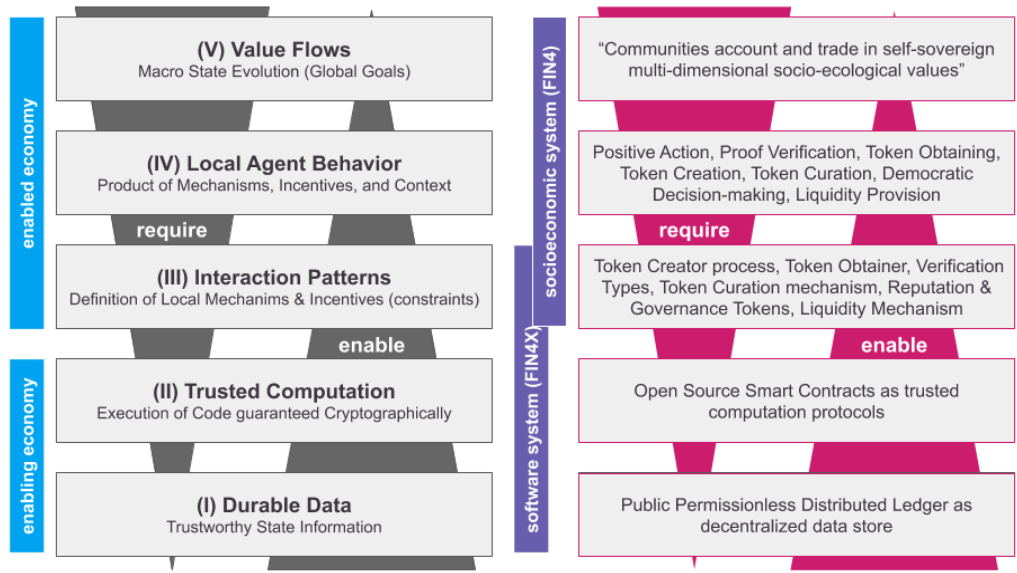

*Figure 1       The socio-ecological Finance 4.0 (FIN4) interacts with and is enabled by a software system (artifact), called FIN4X (adapted from (Dapp, Helbing, and Klauser 2021a, Forthcoming))*

### 4.2.3 System design

We use the model proposed by Zargham (Zargham 2018) to describe the layers of FIN4 as a CES (cf. Figure 1). While cryptoeconomic design takes all five layers into consideration, design principles (see previous section) can only be explicitly formulated and implemented in the software system, i.e. the lower three layers. The software system (called FIN4X for "FIN4 Xplorer") needs to provide durable data storage (layer I), trusted computation on these data (II), and interaction patterns (III) to define what users -- 'agents' from a complex system perspective (Helbing 2012, Chapter 2) -- can and can not do (see Figure 1). Thus, interaction patterns shape the user behavior (IV), locally and across the socioeconomic sphere (i.e., real life). If the mechanisms and incentives were defined adequately, the aggregate behavior of all users should move the system towards the global goals, expressed as value flows (V), envisioned for the entire system. Layer (I) and (II) are facilitated by a smart-contract-enabled blockchain, as illustrated in Section 4.3. While all discussed design principles relate to the FIN4X software system and hence focus on the three bottom layers in Figure 1, the layer in the middle, *interaction patterns (III)*, is shared by both: it is enabled by the software system FIN4X and it is required to allow the socioeconomic system FIN4 to emerge. The core set of interaction patterns in layer (III) consists of the processes creating new tokens, the mechanisms to obtain these tokens and the related mechanism of submitting verifiable proofs for positive actions. To prevent malicious actors from receiving tokens for actions they have not performed, different verification mechanisms are required. Additional interaction patterns are provided on (III) to enable and safeguard adequate local agent behavior: (i) a mechanism to curate tokens enables the user community to identify viable token designs. Token curation is connected to reputation and governance mechanisms (with their own respective tokens): for beneficial behavior in the system, users are rewarded reputation tokens that enable them to participate in the governance of the system. (ii) Also, a mechanism to provide liquidity to a token is provided. For example, a tree token can be backed by another token (e.g. Bitcoin) in addition to having value from the planting of a tree (Ballandies, Dapp, and Pournaras 2021, Forthcoming). The two top layers can not be directly influenced by any software design; they will only emerge if the mechanisms and incentives defined on the interaction pattern layer are adequate: The *Local Agent Behavior* in the FIN4 system should be characterized by the following elements: (i) token creation: Communities and individuals have *permissionless* access to design and deploy cryptoeconomic tokens that incentivize others to perform sustainability actions; (ii) token obtaining: Actors *voluntarily* obtain these tokens by performing actions that the token creators value and verify the performed actions of other actors; (iii) token curation: system participants govern *collectively* a list of accepted tokens, called positive action tokens (PAT) that are supported by the FIN4 system; and (iv) liquidity provision: accepted FIN4 tokens are backed by a *stable* coin that provides an additional source of value to PATs, as illustrated in greater detail in Section 4.3.1. The aggregate behavior of all agents over time is creating *value flows*. In the case of FIN4, the macro state goal is to empower communities to account and trade in self-sovereign multi-dimensional socio-ecological values fostering sustainable action (cf. layer V in Figure 1).

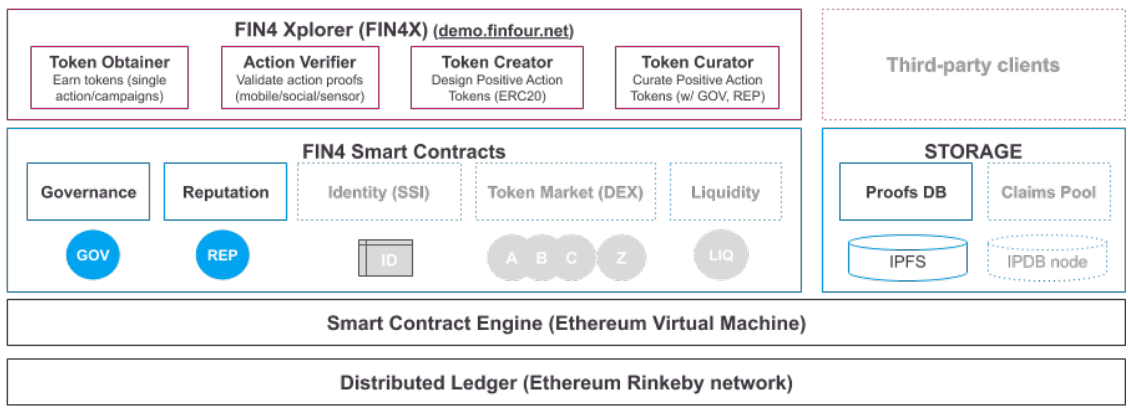

*Figure 2         Software stack of the Finance 4.0 system (adapted from (Dapp, Helbing, and Klauser 2021a, Forthcoming))*

## 4.3    Construction

Based on the identified requirements and design principles (Table 4) a CES is constructed that facilitates the creation, obtaining and curation of cryptoeconomic tokens. The final software artifact[4] is depicted in Figure 2. It consists of smart contracts that are deployed to the Ethereum Rinkeby test network (cf. Layer I, Figure 1). As Ethereum provides a permissionless smart contract engine that is open source, trusted computation is ensured in FIN4 (cf. Layer II, Figure 1'). This choice contributes to support the important values of trustworthiness, resilience and efficiency. Also, the choice facilitates durable data, as Ethereum provides a blockchain-based ledger for immutable storage of user interactions, proof verifications, and tokens. This choice contributes to the values trustworthiness and privacy (i.e., pseudonymity).

Figure 3 illustrates the classification of FIN4X using a taxonomy for DLT systems (Ballandies, Dapp, and Pournaras 2021, Forthcoming), as performed in (Dobler, Ballandies, and Holzwarth 2019) for a supply chain management application. It illustrates the design choices that were taken in the construction of the system.

---

[4]     The artifact can be accessed via: https://demo.finfour.net/ (last accessed November 2020). Please note that the artifact is deployed on the Ethereum Rinkeby test network. The source code of the artifact can be accessed on github: https://github.com/FuturICT2

*Ballandies et al./ Value-sensitive Cryptoeconomic System Design*

| Component | Attribute | Value | | | |
|---|---|---|---|---|---|
| **Distributed Ledger** | Origin | Native | External: Ethereum | Hybrid | |
| | Type | *Blockchain* | DAG | Other | |
| | Address Traceability | *Traceable* | Obfuscatable | | |
| | Turing Completeness | *Yes* | No | | |
| | Storage | *Yes* | No | | |
| **Consensus** | Finality | Deterministic | *Probabilistic* | | |
| | Proof | *PoW* | PoS | Other | |
| | Write Permission | *Public* | Restricted | | |
| | Validate Permission | *Public* | Restricted | | |
| | Fee | *Yes* | No | | |
| **Action** | Actor Permission | *Public* | Restricted | | |
| | Read Permission | *Public* | Restricted | | |
| | Fee | Yes | *No* | | |
| **Token (REP)** | Supply | Capped | *Uncapped* | | |
| | Burn | Yes | *No* | | |
| | Transfereability | Yes | *No* | | |
| | Creation Condition | Action | *Consensus* | Both | |
| | Unconditional Creation | None | Partial | All | |
| | Source of Value | Action | Consen. | DL | *Token* |

*Figure 3        The Finance 4.0 system design in the DLT system taxonomy (Ballandies et al. 2021, Forthcoming)*

### 4.3.1 Token Creator and sources of value

At the core of the FIN4 system lies the Token Creator. It is deployed as a system of smart contracts to the Ethereum Rinkeby network, so it is publicly accessible. Anyone can use the deployed smart contracts[5] to create, in a step-by-step process, new cryptoeconomic tokens to incentivize actions that the individuals or communities value, called Positive Action Tokens (PAT) in FIN4. The process places no limit on the number of communities or tokens. The Token Creator component is informed by the classification of FIN4 in the DLT taxonomy (Figure 3) and is described in detail by Hirschbaeck (2020). Based on the requirements analysis (Section 4.2.1), the FIN4 system is developed as a decentralized application (DApp) on Ethereum. Hence, token attributes that require permission rights[6] on the Ethereum Blockchain can not be set by the token creator. All other attribute values of the DLT taxonomy (Action and Token component in Figure 3) can be set. Thus, the token creator enables the creation of token designs given by the DLT taxonomy. Additional token attributes are implemented: (i) action verifiers that define what proofs users have to provide in order to receive token units for an action; (ii) minting policies that define the amount of minted token units per successful claim; (iii) sources of value that may serve as underlying of a token.

This source of value, which provides value to cryptoeconomic tokens (Ballandies, Dapp, and Pournaras 2021, Forthcoming), is important to make a *multi-dimensional* incentive system work (Kleineberg and Helbing 2016). The creator of a token can combine FIN4 external and internal sources of value. Three internal mechanisms are implemented: The swap enables the exchange of a token unit with another token out of a pool of collected tokens. The mint enables the "burning" of token units that trigger the creation of units of another token (i.e. they will afterwards "disappear"). And the burn facilitates that for each burning of a token unit a token unit of another currency is also burned. As an external source of value, a smart contract can be linked to the token or a free text can be entered illustrating the source of value in greater detail.

---

[5]        The FIN4 Explorer is deployed live and can be accessed here: https://demo.finfour.net/ . Please note that an installation of an Ethereum wallet like metamask is required.
[6]        e.g. a Creation Condition of Consensus (Figure 3), which would link token creation to a participation in the Consensus as it is facilitated in Bitcoin with the block reward.

### 4.3.2 Token Obtainer and Action Verifier

The FIN4 system links sustainable actions in the real world to the creation of new token units as an action reward (cf. (Ballandies, Dapp, and Pournaras 2021, Forthcoming). The token creator defines actions which warrant a token unit as well as a list of verifiers to check proof of these required actions. Currently 14 verifiers are implemented in the FIN4 systems. In order to increase the resilience against malicious actors who attempt to obtain tokens for non-performed actions (see Section 4.2.3), a user can combine several of the verifiers and also add new verifiers if the one provided are not sufficient for a specific use case. Inclusiveness and autonomy is facilitated by involving humans in the verification process and thus reducing the dependence on algorithmic mechanisms. Verifications are logged permanently in the distributed ledgers to facilitate the transparency and accountability of the system. Implementation details of the Action Verifier mechanisms are provided in (Rachwan and Chodyko 2021).

### 4.3.3 Token Curator and reputation

A Token Curated Registry (TCR) is a mechanism that utilizes "economic incentives for rational stakeholders to curate a list" of items (Soima 2019, 9). Such a mechanism is utilized as a governance instrument in FIN4 for curating positive action tokens. The public and collaborative curation process utilizes hidden votes, staking, incentivization and other game theoretical mechanisms to create a list of curated tokens that align with the values of the systems stakeholders and that enable the incentivization of sustainability-based actions. Implementational details of the TCR and its mechanisms are provided by Soima (2019). To participate in token curation, FIN4 governance tokens (GOV) are required that can be obtained either from other users by delegation or by claiming them after reaching a specific level of reputation (REP), a second non-transferable FIN4 token. The REP token properties are illustrated in Figure 3. Currently, REP can be earned in the FIN4 system by creating tokens with the token creator (Section 4.3.1) or by successfully claiming another token with the token obtainer (Section 4.3.2). When owning a sufficient amount of GOV tokens, users can propose tokens to be added to or be removed from the curated list and participate in the voting that decides on such proposals.

### 4.4 Use of Software

The created software artifact was used in two use case scenarios. At a forum focus group (Section 3.3), 50 participants created and interacted with 11 cryptoeconomic tokens of which 7 were created by participants, the rest by the research team. These tokens utilized 7 verifiers (Section 4.3.3). In total 139 claims were submitted by utilizing the software and 72 token units were created by passing the verifiers. Another use case was conducted with a NGO that engages in the future of work. The NGO was characterized by different teams, which owned and shared various resources. A multi-dimensional token system was designed that encouraged the utilization and exchange of these resources. During three Webinars, eight members of the NGO interacted with the software and responded to a survey. The results of the survey are illustrated in Table 3. The majority of users perceive FIN4 tokens to improve the capturing of values when compared to money. In particular they state that FIN4 tokens reduce the need to use money. Despite interacting with the system via a wallet that is controlled by an own private key, half of the users think that other people can control their cryptoeconomic tokens.

| Question | strongly disagree | disagree | undecided | agree | strongly agree |
|---|---|---|---|---|---|
| FIN4 tokens express more/ different values compared to money. | 0 | 0 | 3 | 3 | 2 |
| I trust that only I have access to my FIN4 tokens. | 0 | 1 | 3 | 3 | 1 |
| FIN4 tokens reduce the need to use money | 0 | 1 | 2 | 3 | 2 |

*Table 3      Responses by the participants on their perception of FIN4 tokens.*

# 5  Summary of findings and implications

| Value | Design principle | Implementation/ Realisation in FIN4 system |
|---|---|---|
| **Resilience (Diversity)** | Decentralization | More than one community implements their values. |
| | Multi-dimensional | Universal Token creator: More than one token are created and designed according to communities values. |
| **Autonomy** | Permissionless | Bottom-up Token creation and governance; no client-server architecture (DApp via smart contracts). |
| | Opt-in | Participation is voluntary. |
| **Sustainability/ human welfare** | Internalisation of negative externalities | Sustainability action-based value capturing mechanism of tokens via source of value and creation condition. |
| **Privacy** | Pseudonymous | No registration is required: The identity of users is build via reputation. |
| **Ownership/ inclusiveness (Participation)** | Direct democracy | Curation of positive action tokens is governed by a token curated registry. |
| | Human participation | Human involvement in creation and obtaining of tokens and verification. |
| **Efficiency** | Automation | Simplified token creation, obtaining and verification via smart contracts. |
| **Trust(worthiness)** | Transparency | Public smart contracts illustrating inner mechanics of systems functioning (white box). |
| | Accountability/ Auditability | An immutable and public transaction log facilitates the transparency of performed actions in the system. |

*Table 4      Values stakeholders of a socio-ecological financial system hold, the design principles in which these got translated and how they were implemented in the Finance 4.0 system*

By combining a value-sensitive design with a Design Science Research Methodology, the following results were obtained: (i) 7 values were identified that stakeholders of a socio-ecological financial system find important (Column 1 in Table 4). (ii) Based on the identified values, 11 design principles were derived (Column 2 in Table 4). (iii) A software artifact was created that implements the 11 design principles (Column 3 in Table 4) and thus demonstrates the feasibility of the design. (iv) The high participation and interaction with the software artifact at workshops, forums and hackathons indicate the relevance and usability of the created software instance. (v) A preliminary study indicates that the created software instance makes users perceive cryptoeconomic tokens as different from usual money.

The developed solution considers insights from complexity science and facilitates a value-sensitive constraint configurator that is democratically governed and which enables the creation of cryptoeconomic systems (CES) in the form of multi-dimensional incentive systems. It can be utilized to construct systems influencing human behavior towards sustainability. Nevertheless, because, local agent behavior and the value flows of the system only emerge when applying the software artifact in real-world settings (Section 4.2), these cannot be fully anticipated at design time. The evaluation of these behaviors and value flows could not be evaluated with the conducted experiment. It requires further longitudinal studies with larger participant fields.

IS research can leverage the developed methodology to design value-sensitive and validated CES. Moreover, it can apply the constraint configurator to research the impact of cryptoeconomic tokens on human behavior. Both, researchers and practitioners can utilize the introduced insights from complexity science and best-practices for comprehensively improving and illustrating the design of their cryptoeconomic systems (Section 4.2 and 4.3). Finally, the latter may utilize the found design principles of a socio-ecological financial system and their instantiations to construct a money system that aligns with the values of its stakeholders. Recent discussions of a digital euro and central bank digital currencies indicate a need for an update of our current system (Carapella and Flemming 2020; Governing Council ECB 2020; Group of Thirty 2020; IMF 2020).

# 6 Conclusion

In conclusion, we can answer the three research questions (Section 1) as follows. First, IS can facilitate the construction of crpyoteconomics systems (CES) that use tokens to represent positive actions and thus incentivize human behavior toward socio-ecological goals (cf. RQ1 and Section 4.1). Second, a methodology to design, implement and validate such CES needs to align stakeholder values with the interaction patterns enabled by the software artifact. This can be achieved by complementing general design science research methods with specific methods for evaluation patterns and value-sensitive design (cf. RQ2 and Section 3). Third, To enable a socio-ecological finance system using a CES, it is important to translate stakeholder values into key design principles from which to derive interaction patterns that evoke the desired local agent behavior over time (cf. RQ3 and Sections 4.2 and 4.3). The Finance 4.0 system is an example of a socio-ecological financial system instantiated as a CES. The design and construction of this system are comprehensively illustrated and documented utilizing state-of-the art cryptoeconomic models, architectures and taxonomies. Preliminary results validate the artifact's relevance and usability in real-world environments and its appreciation by stakeholders (Section 4.4).

The results point to various avenues for future research. First, because the macro state of any CES is an emergent property of local agent behavior, the exact outcome cannot be anticipated at the design phase. To mitigate the risk of reaching unintended system states, these behaviors should be evaluated in simulations and longitudinal experiments with larger participant fields. This could provide insights into the robustness of interactions patterns such as the FIN4 token curated registry. Second, preliminary study results indicate that users perceive cryptoeconomic tokens as different from money which may have an effect on the impact of these tokens on human behavior and motivation. Third, the FIN4 system can act as a constraint configurator in behavioral experiments, thus serve as a tool to test the effects of multi-dimensional token-based incentive systems on human behavior in real-live settings.

# 7 Acknowledgements

The authors acknowledge funding by the Swiss National Foundation for EU FLAG ERA project FuturICT2.eu under grant number 170226. The authors would like to thank Qusai Jouda, who created the very first version of the FIN4 code base, as well as Dian Balta and participants of the "Distributed Ledger Technology for Public Sector Innovation" course at TU Munich, namely Piotr Chodyko, Sangeeta Joseph, Leon Kobinger, John Rachwan, Moritz Schindelmann, Simon Zachau, and Ling Zhu, who have also contributed to the code base besides ETH Zurich. Further thanks go to students who wrote their thesis or semester projects about Finance 4.0: Leonie Flückiger, Gabriel Hirschbaeck, Max Rößner, Kriti Shreshtha, and Sergiu Soima. Last but not least, we acknowledge Anabele-Linda Pardi, Stefan Klauser, Thorben Funke, Evangelos Pournaras, Alexander Stein, and Magnus Wuttke for their contributions and helpful comments.